\documentstyle[a4]{article}

\begin{document}

\newcommand{\bib}{\bibitem}
\newcommand{\er}{\end{eqnarray}}
\newcommand{\br}{\begin{eqnarray}}
\newcommand{\be}{\begin{equation}}
\newcommand{\ee}{\end{equation}}
\newcommand{\epe}{\end{equation}}
\newcommand{\bea}{\begin{eqnarray}}
\newcommand{\eea}{\end{eqnarray}}
\newcommand{\ba}{\begin{eqnarray}}
\newcommand{\ea}{\end{eqnarray}}
\newcommand{\epa}{\end{eqnarray}}
\newcommand{\ar}{\rightarrow}

\def\r{\rho}
\def\D{\Delta}
\def\R{I\!\!R}
\def\l{\lambda}
\def\D{\Delta}
\def\d{\delta}
\def\T{\tilde{T}}
\def\k{\kappa}
\def\t{\tau}
\def\f{\phi}
\def\p{\psi}
\def\z{\zeta}
\def\ep{\epsilon}
\def\hx{\widehat{\xi}}
\def\na{\nabla}
\begin{center}

{\bf Hodge-type self(antiself)-duality for general p-form fields in
arbitrary dimensions. }

\vspace{1.3cm} M. Botta Cantcheff\footnote{e-mail: botta@cbpf.br}

\vspace{3mm} Centro Brasileiro de Pesquisas Fisicas (CBPF)

Departamento de Teoria de Campos e Particulas (DCP)

Rua Dr. Xavier Sigaud, 150 - Urca

22290-180 - Rio de Janeiro - RJ - Brazil.
\end{center}

\begin{abstract}
It is often claimed \cite{PST1} that the (Hodge type) duality operation is
defined only in even dimensional
spacetimes and that self-duality is further restricted to twice-odd dimensional
spacetime theories.
The purpose of this paper is to extend the notion of both duality symmetry
as well as self-duality.

By considering tensorial doublets, we introduce a novel
well-defined notion
of self-duality based on a duality Hodge-type operation in
arbitrary dimension and for any rank of these tensors.
 Thus, a generalized Self-Dual Action is defined such that
 equations of motion are the claimed generalized self-duality
 relations. We observe in addition,
 that taking the proper limit on the parameters
of this action, it always  provides us with a
 master-action, which interpolates
 models well-studied in physics;  by considering a particular limit,
 we find an action which describes
 an interesting type of relation, referred to as semi-self-duality,
 which results to be the parent action between Maxwell-type actions.

 Finally, we apply these ideas to construct manifest Hodge-type
 self-dual solutions in a (2+1)-dimensional
  version of the Maxwell's theory.

\end{abstract}

\section{Introduction}

The purpose of this paper is to explore a certain freedom in the definition
of the duality operation in order to construct self and anti-self-dual
models\cite{review}

This issue is nowadays well-motivated \cite{witten}. The role of
duality in the investigation of physical systems is by now well-appreciated.
 This duality symmetry, that is fundamental in the
current understanding of quantum field theory, statistical
mechanics and string theory, is a general concept relating
physical quantities in different regions of the parameter space.
It relates a model in a strong coupling regime to a distinct one
in a weak coupling regime, providing a valuable mechanism for
investigating strongly interacting models.

 One currently defines the Hodge-Duality (HD) operation by the
contraction with the totally antisymmetric $\epsilon$-symbol. There
is also an extension of this operation used for instance, in 2+1-dimensions
, which is basically a functional curl (rotational operator):
\be
\label{1}
\mbox{}^{*} f_\mu =\frac{\chi}{m}\,
\epsilon_{\mu\nu\lambda}\,\partial^\nu f^\lambda\,,
\ee
 where $m$ is a constant to render the
 $\mbox{}^{*}$-operation
 dimensionless. We call this Differential Hodge Duality (DHD),
and in all cases, we name {\it self(anti-self)-duality}, when the relations
$\mbox{}^{*} f = \pm f $ are (respectively) satisfied.

The so-called Self-Dual Model \cite{TPvN,annals,DJ} is given by
the following action,

\be
\label{180}
 S(f)= \int\, d^3x\:\Bigg(\frac {\chi }{2m}\,
\epsilon_{\mu\nu\lambda}\,f^\mu\,\partial^\nu f^\lambda -
\frac{1}{2}\, f_\mu f^\mu \Bigg)\,.
\ee

The equation of motion is the self-duality relation:
\be
\label{190}
f_\mu =\frac{\chi}{m}\,
\epsilon_{\mu\nu\lambda}\,\partial^\nu f^\lambda\,,
\ee
This model is claimed to be chiral, and the chirality $\chi=\pm 1$ results
 defined precisely from this self-duality.

  A (Hodge-type) well-defined self ( and anti-self )-duality between tensor
   fields with different ranks is lacking
  in the literature. This concept
  is important by itself;
this can be applied, as we have suggested, in a formal
generalization of the notion
 of chirality.

 The possibility of defining self-duality and non merely duality between two systems
is critically important also in topological configurations; in particular,
it can provide us with a generalized notion of instantons.

The question we would like to address regards the possibility of
imposing self-duality in general dimensions. This
 will be answered affirmatively, as will be shown below.

 Let us describe better the problem we are concerned with:

Consider a general $q$-form

\be \label{a05} F_{\mu_1 \cdots \mu_q} . \ee The Hodge-dual field
is then defined as,

\be \label{a10} \mbox{}^* F^{\mu_{q+1}\cdots \mu_d} = \frac 1{q!} \;
\epsilon^{\mu_1\cdots \mu_{d}} F_{\mu_{1}\cdots \mu_{q}} \ee
Note that, in principle, only in $d=2q$ dimensions one could
define self-duality, since
 the $q$-form field $F$ would be of the same rank as its dual.

This is the type of duality present in Maxwell-type theories, where $F$
is a exact differential, i.e $F=df$ for a $q-1$-form field $f$.

In this context, the field equation and Bianchi identity
 for a source free field are

\ba
\label{a15}
0&=& \partial_{\mu_1}F^{\mu_1\cdots \mu_q} \nonumber\\
0&=& \partial_{\mu_1}\mbox{}^*F^{\mu_1\cdots \mu_q} \ea

The field equation and the Bianchi identity
are of the same form so that the duality transformation $F\leftrightarrow \mbox{}^* F$
is a symmetry but in general, is not present at level of the tensor-fields.

The dependence with dimensionality appears to be crucial.
%There is another question where the dimensionality appears to be crucial for defining self-duality

 It is well-known the problem of defining the Hodge duality for all
dimensions; for instance, in Lorentzian four-dimensional spacetime,
 the main obstruction to
self-duality comes from the relation of
double-dualization for a rank-two tensor

\be \label{Q10} \mbox{}^{**} F = (-1)^{s}\; F \ee 
where $s$ is the
signature of the Minkowski metric.  For the case of the Lorentzian metric, where $s$ is an
odd number, the self-duality concept seems inconsistent with the
double dualization operation due to the minus sign in (\ref{Q10}).
This problem remains for dimensionality $d=4m$ ($m\in Z_+$)
\cite{kn:DGHT} \cite{W} \cite{ban-w}, in contrast, it is absent for $d=4m-2$.
Thus, self-duality is claimed to be well defined (only) in such
dimensionality. The self-duality present in $d=4m-2$ dimensions has
attracted much attention because it seems to play an important
role in many theoretical models\cite{EK}. The possibility we are
discussing in this work should allow us to extend these
applications.

This work is organized according to the following outline:

     In Section 2, these difficulties and its resolution are clarified,
      the notion of duality is generalized, allowing to construct a generalized manifestly
      self-dual action; next, in Section 3, we discuss the relation of this action
      with the so-called parent action, which interpolates between several couples of
      (dual) equivalent models\cite{suecos}. Finally, in Section 4, as an application, we build
      Hodge duality in 2+1-dimensions in a similar fashion with the Maxwell's theory
       of the electromagnetism. This construction is new.

\section{Generalization of HD (-DHD) and self(anti-self)-duality.}

In order to remove the obstruction for consistency between duality
and self-duality posed by the presence of the minus sign in
(\ref{Q10}), we start by exploring the existence of an ambiguity
in the double dual operation that comes around from the fact that
the duality operation is a mapping from the space of $d/2 $ forms
to its co-space, $*\, :\Lambda_{(0,d/2)}\to \Lambda_{(d/2,0)}$, so
that the inverse mapping is not automatically defined. This leaves
some room for distinct alternatives with interesting consequences.
First, let us recall that (\ref{Q10}) has led to the prejudice
that the (Abelian) Maxwell theory would not possess manifest
self-duality solutions. The resolution for this obstacle came with
the recognition of an internal two-dimensional structure hidden in
the space of potentials. Transformations in this internal duality
space extends the self-duality concept to this case and is
currently known under the names of Schwarz and Sen\cite{SS}, but
this deep unifying concept has also been appreciated by
others\cite{Z,DT}.  The actions obtained corresponds to self-dual
and anti-self-dual representation of a given theory and make use
of the internal space concept.  The duality operation is now
defined to include the internal (two dimensional) index $(\alpha, \beta)$ in the
fashion

\be
\label{Q20} \hat F^\alpha =e^{\alpha\beta}\; \mbox{}^{*}F^\beta \ee

where the $2 \times 2$ matrix, $e$, depends on the signature and dimension 
of the spacetime in the form:

\be
\label{a20}
e^{\alpha\beta}= \cases{\sigma_1^{\alpha\beta},&if $d=4m-2$\cr
	\ep^{\alpha\beta},&if $d=4m$.\cr}
\ee

with $\sigma_1^{\alpha\beta}$ being the first of the Pauli matrices
and $\epsilon^{\alpha\beta}$ is the totally antisymmetric $2 \times 2$ matrix with $\ep^{1,2}=1$.
 
  The double duality operation 

\be
\label{Q30} \hat {\hat F} = F \ee generalizes (\ref{Q10}) to allow
consistency with self-duality. In \cite{RW} we show how the prescription
(\ref{Q20}) works in the construction of self-dual Maxwell actions.

 Most of the usual discussion about duality transformations as a symmetry for the
actions and the existence of self-duality are based on
these concepts.

\vspace{0.7cm}

Remark: Up to now, this structure has always been defined only for
tensorial objects where the field has the same tensorial rank that
its corresponding dual.

\vspace{0.7cm}

Now, we generalize further these ideas, introducing
more general doublets. This concept is new
and constitutes our main contribution.

 The matters described above can be avoided by defining Hodge duality
(HD) and differential Hodge duality (DHD) for tensorial doublets
in a matricial form. Let a p-form (a totally antisymmetric tensor
type $(0;p)$) on a $d$-dimensional space-time with signature $s$ \footnote{i.e, 
this is the number of minuses occurring in the metric},
$f_{\mu_1,.....\mu_p}$, one natural partner shall be any
$((d-p);0)$-tensor, $g_{\mu_1,.....\mu_{d-p}}$. We build the tensor
doublet $F:=(f,g)$.

{\bf Hodge Duality: }

Let us define now the {\it generalized} Hodge-operation \footnote{ For notational convenience, in order to avoid
explicit indices, in the following lines we forget thet
 the duality operation must be defined as a map from the field-space into its co-space.}
($()\mbox{}^{*}$) for this object by means of
\be
\mbox{}^{*}F:=(\mbox{}^{*}g \, , \,S \mbox{}^{*} f  ),
\ee
 where
\be
(\mbox{}^{*}f )^{\mu_{p+1}\cdots \mu_{d}} = \frac 1{p!} \;
\epsilon^{\mu_1\cdots \mu_{d}} f_{\mu_{1}\cdots \mu_{p}} \, ,
\ee
and $S$ is a number defined by the double dualization operation:
\be
\mbox{}^{*}(\mbox{}^{*}f) = S f \,.
\ee
This depends on the signature $(s)$ and spacetime dimension in the form $S=(-1)^{s+p[d-p]}$,
 which clearly includes the case $p=d/2$ described above.

Notice that this Hodge-type self (anti-self)-duality {\it is well-defined},
since 
 \be
 \mbox{}^{*}F=\pm F\, ,
\ee
is consistent with the double dualization requirement, $\mbox{}^{*}(\, \mbox{}^{*}\, F )=F$.

{\bf Differential Hodge Duality: }

Furthermore, DHD can be also generalized in this fashion: let us consider
doublets $F=(f_{\mu_1,.....\mu_p};g_{\mu_1,.....\mu_{d-p-1}})$ then,
in terms of forms, DHD is defined as

\be
\label{DHD}
 \mbox{}^{\star}  F \equiv K^{-1} ~\mbox{}^{*}d F\, , 
\ee
where $d(f,g)=(df , dg)$ and the matrix $K=diag[k_f ; k_g]$, that is
 taken to be diagonal for simplicity, is introduced for dimensional
 reasons \footnote{ i.e $k_f$ and $k_g$ must have dimension of mass.} .

Thus,  self (anti-self)-duality  is also 
well-defined in this case, since the relations 
 \be
\label{sdrel}
 \mbox{}^{\star}  F =\pm F  
\ee
may be consistent with the double dualisation
requirement, $\mbox{}^{\star} ( \mbox{}^{\star}  F ) = F$

The self-duality relation in this case reads
\be
\label{sdrel}
 F = ({\chi}K^{-1})~\mbox{}^{*} dF   ,
\ee

where $\chi=\pm 1$ .
As it can be trivially verified, the consistency of self-duality requires
 that $F$ satisfies the Proca equation with mass $m=\sqrt{k_f k_g }$. In fact,
 applying once more
  the operator $\mbox{}^{\star} $ to (\ref{sdrel}), we have:
\be [\partial ^2 + m^2] F =0 .\ee

The next step is to obtain an action which expresses self-duality in
 this generalized sense.
Then, we write down this one, which is a {\it generalized} Self-Dual Model (GSDM) in $d$-dimensions
\footnote{The doublet internal product of pairs "${\bf .}$"
is naturally given by $(f,g){\bf .}(f' , g') \equiv f_{\mu_{1}\cdots \mu_{p}}\,f'^{\mu_{1}\cdots \mu_{p}}\,
+ g_{\mu_{1}\cdots \mu_{q}}\,g'^{\mu_{1}\cdots \mu_{q}}$, where $f ,\,f'$ are
 $p$-forms, while $g ,\,g'$ are any $q$-forms.} :

\be
 {\cal S}_{GSD}[ F]= \int\,d^{d}x\, \left( \frac{\chi}{2}  F ~{\bf . }~ \mbox{}^{*} d F
  +  \frac{1}{2}  F~K~ F\right)~~~~~~.
\label{GSDM} \ee 
This is the central object of this work. 
It is straightforward to verify that the equations of motion are precisely
 the self-duality relations (\ref{sdrel}).
 Notice that this action looks like SD-action in three dimensions, Eq. (\ref{180}), 
and here is its main importance since the structure of these theories (in $2+1$) can
naturally be extended to arbitrary dimensions. Results based on this 
issue, in the context of topologically massive
theories  \cite{tmdob} and bosonization in general dimensions \cite{bos}, are being reported
 elsewhere.

Furthermore, as we show below, one can
obtain, by taking or not, appropriate limits of the 
 constants $(k_f ;k_g)$,
this corresponds to different Parent Actions, and so describing different
 dualities between models whose fields are
precisely the components of the general doublet $F$ ( see \cite{suecos} and references therein ).

\section{GSDM and Parent Actions.}

As it has been motivated, in this section we argue that the parent actions
describing duality between physically relevant models can be obtained
from the GSDM-action by a different fixing of the tensor-doublet, the duality operation,
and the coupling matrix $K$.

A very nice structure of dualities arises from this elegant
analysis. Depending on the $K$-parameters we take,
interesting consequences are obtained. For instance, if
$k_g=k_f=0$, we get a topological theory; If in contrast, $K$ is
non-singular, this is the master action between two Proca models
for both $f$ and $g$. We show this later.

The main result is
 obtained by considering only $k_g=0$. This model reflects a sort
  of semi-self-duality as we will show,
and provides us with the master action {\it in general}.
 We will illustrate this by means of a simple example
and in the following subsection we shall prove this in general.

\subsection{Example: Scalar-Tensor Duality}
 This is an example of duality between two systems of different
 tensorial ranks, where the relations of
self-duality cannot be written. According to our prescription, we shall show that
 self-duality could be defined for this system,
via an action of the type (\ref{GSDM}). With a proper limit to obtain a singular $K$-matrix,
 we recover the familiar parent actions
of this problem. This example was discussed in a very illuminating way by Hjelmeland and 
Lindstr\"om\cite{suecos}:

Consider the action for a massless free Klein-Gordon
 field $\phi$ in $d=4$
\begin{equation}
S(\phi)=\frac{1}{2}\int d^{4}x\partial_{\mu}\phi
\partial^{\mu}\phi, \label{action-fi}
\end{equation}

the field equation and the Bianchi identities for the free Klein-Gordon field are respectively:
\begin{eqnarray}
\partial_{\mu}\partial^{\mu}\phi
&=&0 \\ \nonumber \\
{\partial}_{\mu}^{~~*}\phi^{\mu\nu\rho}(\phi) &=&0
\end{eqnarray}
where ${~}^{*}\phi^{\mu\nu\rho}\equiv\epsilon^{\mu\nu\rho\sigma}
\partial_{\sigma}\phi .$

On the other hand, the action for a free massless
 two-form field $A_{\mu\nu}$ is
\begin{equation}
S(A)=\frac{1}{3!}\int
d^{4}x\partial_{[\mu}A_{\nu\rho]}\partial^{[\mu}A^{\nu\rho]};
\label{action-A}.
\end{equation}
 Now, we write down the Bianchi identity and the field equations, respectively:
\begin{eqnarray}
{\partial}_{\mu}^{~~*}A^{\mu}
&=&0 \label{field-eq1} \\ \nonumber \\
\partial_{\mu}\partial^{[\mu} A^{\nu\rho]}
&=&0
\end{eqnarray}
where ${~}^{*}A^{\mu}\equiv\frac{1}{3!}
\epsilon^{\mu\nu\rho\sigma}\partial_\mu A_{\nu\rho}$.

The main point is that the field equation for
 the free Klein-Gordon field looks like the Bianchi
  identity for the free anti-symmetric field, and vice-versa.
   The change from
  one description to the other interchanges the roles of the field
  equations and the Bianchi identities.
At the classical level, they are two models representing the same
physics. To show this explicitly, it
    is currently introduced
       the so-called {\it parent action}:
\begin{equation}
S_p({F_{\mu\nu\rho},\phi})= \frac{1}{3!}\int
d^{4}x\left(F_{\mu\nu\rho}F^{\mu\nu\rho}+
\sqrt{2}\phi\partial_{\mu}^{~*}F^{\mu}\right), \label{parent1}
\end{equation}
where $\phi$ and $F_{\mu\nu\rho}$
 are independent field.
  Varying this action with respect to
  $\phi$ gives:
\begin{equation}
 0=\epsilon^{\mu\nu\rho\sigma}\partial_{\mu}F_{\nu\rho\sigma}.
\end{equation}
Hence, there exists a two-form field, $A_{\rho\sigma}$, such that
$F_{\nu\rho\sigma}=\partial_{[\nu}A_{\rho\sigma]}$. Putting this
back into the action (\ref{parent1}),  we recover the action
(\ref{action-A}). We have thus shown that (\ref{parent1}) is
equivalent to (\ref{action-A}). Substitution of solutions of field
equations into the parent action requires that consistency at
level of field equations be verified; however, in this case, this
is not a serious matter and the (on-shell) equivalence between the
parent action and (\ref{action-A}) is verified.

Now, in order to show that (\ref{parent1}) is also equivalent to
(\ref{action-fi}),
 we
vary (\ref{parent1}) with respect to $F_{\mu\nu\rho}$, thus:
\begin{equation}
F^{\mu\nu\rho}=-\frac{1}{\sqrt{2}}\epsilon^{\mu\nu\rho\kappa}\partial_{\kappa}\phi
.
\end{equation}
Replacing this into $S_p(F,\phi)$ we obtain
\begin{eqnarray}
 S_p(F[\phi],\phi)= S(\phi) = \frac{1}{2}\int
d^{4}x\partial_{\mu}\phi\partial^{\mu}\phi .
\end{eqnarray}

We have shown, using the parent action,
 that $S(\phi)$ and $S(A)$ are dual to each other; the two actions
  describe the same physical system, but the physical
   representation is given using different fields. A remarkable feature
    of this construction is that the field equations and the Bianchi identities
     are exchanged.

     According to the doublets structure presented in the preceding
     section, we define the doublet ${\cal F}=(\phi, F )$, and
      self-duality relations for this system can be written as two simultaneous
     equations:

\begin{equation}
\label{s1} F^{\mu\nu\rho}=
-\frac{1}{k_F}\epsilon^{\mu\nu\rho\kappa}\partial_{\kappa}\phi ,
\end{equation}

    \begin{equation}
    \label{s2}
\phi=
\frac{1}{k_\phi}\epsilon^{\mu\nu\rho\kappa}\partial_{\kappa}F_{\mu\nu\rho} ,
\end{equation}

    which may be derived of an theory with the form GSDM (Equation ({\ref{GSDM})),
    \be
\label{lagdob1}
    S_{\cal{F}}({\cal F})= \int
d^{4}x {\cal F} [. \mbox{}^{*}d  \,+ \,K\,] {\cal F}
    \ee
    where
$K=diag(k_F;k_\phi)$. However, if we take $k_\phi=0$ and $k_F=\sqrt{2}$, we clearly obtain
relations,
  instead of
(\ref{s1}) and (\ref{s2}), which can be interpreted as a sort of semi-self-duality.
In fact, the action (\ref{lagdob1}) dictates the form of the parent action:
\be
S_{\cal F}({\cal F})=  \int
d^{4}x\left(F_{\mu\nu\rho}\sqrt{2}F^{\mu\nu\rho} +
\phi\epsilon^{\mu\nu\rho\alpha}\partial_{\mu}F_{\nu\rho\alpha} -
 F_{\nu\rho\alpha}\epsilon^{\mu\nu\rho\alpha}\partial_{\mu}\phi\right) =
  \frac{3}{\sqrt{2}} S_p({F_{\mu\nu\rho},\phi}) + \mbox{Total divergence}.
\ee
This structure is illustrated in the Fig.\ref{dualityA} below.

\begin{figure}
\begin{center}
\setlength{\unitlength}{1mm}
\begin{picture}(100,70)
\put(48,58){$S_{\cal F}$}
\put(46,57){\vector(-2,-3){23}}\put(54,57){\vector(2,-3){23}}
\put(22,40){$\delta
F_{\mu\nu\rho}$}\put(68,40){$\delta\phi$}
\put(18,17){$S(\phi)$}\put(78,17){$S(A)$}
\put(20,10){\vector(0,1){5}}\put(80,10){\vector(0,1){5}}
\put(20,10){\line(1,0){20}}\put(60,10){\line(1,0){20}}
\put(43,9){\bf{duality}}
\end{picture}
\caption{}
\label{dualityA}
\end{center}
\end{figure}

On the other hand, we obtain remarkably the parent action (\ref{parent1}).
Parent actions are not unique, as well as doublets in a given dimension.

 Below,
 we understand
that, in dimension four, another doublet can be chosen ${\cal G}
=(A_{\nu\rho},f_\mu)$,
 resulting in the same duality.

 Thus, taking $k_f = \frac{1}{\sqrt{2}} ;k_A=0$,
the other parent action, which also shows that $S_{\phi}$ and $S_{A}$ are dual to
 one
 another, reads as follows:
\ba
\label{par2}
S_{\cal G}({\cal G})&=& \int d^{4}x {\cal G} [. \mbox{}^{*}d  \,+ \,K \,]  {\cal G} \nonumber\\
 &=&\int d^{4}x \left({\cal G}\,. \mbox{}^{*}d    {\cal G} +  \frac{1}{\sqrt{2}}f_{\mu}f^{\mu}\right) \nonumber\\
&=&\frac{1}{\sqrt{2}}\int d^{4}x\left(\frac{1}{2}f_{\mu}f^{\mu}+
\sqrt{2}A_{\mu\nu}\epsilon^{\rho\mu\nu\sigma}\partial_{\rho}f_{\sigma}\right) + \mbox{Total divergence},
\ea

where $f$ and $A$ are independent fields.

Varying this action with respect to $A_{\mu\nu}$, we obtain
\begin{equation}
0=\epsilon^{\rho\mu\nu\sigma}\partial_{\rho}f_{\sigma}.
\end{equation}
Again, this implies that there exists a scalar field $\phi$ (at
least locally) such that $f_\mu=\partial_\mu \phi$

Replacing this expression in (\ref{par2}), we have: \be S_{\cal{G}}
= \frac{1}{\sqrt{2}} S(\phi)\ee

 Varying now the parent action with respect to $f_{\mu}$, we obtain
\begin{equation}
f^{\mu}=-\sqrt{2}\epsilon^{\mu\nu\rho\sigma}\partial_{\nu}A_{\rho\sigma}.
\end{equation}
Plugging this back into $S_{F,A}$, we get

\begin{equation}
S_{\cal G}(f[A],A)=\frac{1}{\sqrt{2}} S(A)=\frac{1}{3!\sqrt{2}}\int
d^{4}x \partial_{[\mu}A_{\nu\rho]}\partial^{[\mu}A^{\nu\rho]}.
\end{equation}

 This dual equivalence is shown in Fig.\ref{dualityB}.
\begin{figure}
\begin{center}
\setlength{\unitlength}{1mm}
\begin{picture}(100,70)
\put(48,58){$S_{\cal G}$}
\put(46,57){\vector(-2,-3){23}}\put(54,57){\vector(2,-3){23}}
\put(25,40){$\delta A_{\mu\nu}$}\put(68,40){$\delta
f_{\mu}$}
\put(18,17){$S(\phi)$}\put(78,17){$S(A)$}
\put(20,10){\vector(0,1){5}}\put(80,10){\vector(0,1){5}}
\put(20,10){\line(1,0){20}}\put(60,10){\line(1,0){20}}
\put(43,9){\bf{duality}}
\end{picture}
\caption{ }
\label{dualityB}
\end{center}
\end{figure}

We refer to these models as Maxwell-type, because their actions have the
 form $S(f) \sim \int (df)^2$.
 As we shall see below in detail, semi-self-dual actions, with duality DHD, describe dualities
  between this type of
 theories in general.

 Another important duality for the doublet ${\cal G}$ shall be mentioned.
The action (\ref{par2}), with $K$ non-singular, namely
\be
\label{dob2k}
S({\cal G})= \int d^{4}x \left( {m\over 6} f_{\sigma}\epsilon^{\sigma\rho\mu\nu}\partial_{[ \rho}A_{\mu\nu ]} +
{m^2 \over 2} f_{\sigma}f^{\sigma} - {m^2 \over 4} A_{\mu\nu}A^{\mu\nu} \right),
\ee
also reveals the connection between the Proca model,

\be
S_{Proca} = \int d^{4}x \left( -{1 \over 2} \partial_{[ \rho} f_{\mu ]}\partial^{[ \rho} f^{\mu ]} + 
m^2  f_{\mu}f^{\mu} \right),
\ee
and the Kalb-Ramond model \cite{ban95}
whose action is given by,
\be
S_{K-R} = \int d^{4}x \left( {1 \over 6} \partial_{[ \rho}A_{\mu\nu ]}\partial^{[ \rho}A^{\mu\nu ]} - 
{1 \over 6}  A_{\mu\nu}A^{\mu\nu} \right).
\ee
Notice finally, that the action (\ref{dob2k}) is
 manifestly {\it self-dual} for the doublet field, ${\cal G}$.

\subsection{Semi-self-duality and Maxwell-type theories.}

A system will be said to be Semi-Self-Dual iff

\be
\mbox{}^{*}d   F= \chi K P_i F,
\ee

where $P_i ; i=1,2.$ represents the two projectors on the internal
 two-dimensional space. These relations can be derived from a GSDM
  action when the mass matrix is singular in the form $k_{j \neq i}=0$.
Consider the doublet $(f,g)$, thus, this action is \footnote {We take for instance, $k_1=k_f=0$.}:

\be
L_p = F\,. \mbox{}^{*}d   F + k_g g^2 .\label{pargen}
\ee
  We shall show here that this action constitutes a parent action which
   interpolates two Maxwell type theories in general. For simplicity, let's take $k_g=1$

The equations of motions for (\ref{pargen}) will be, if we vary with respect to $g$:
\be
\mbox{}^* d f= g , \label{semi1}
\ee
 Integrating the parent action by parts leads to:
\be
L_p = g \, \mbox{}^* d f  + S f \, \mbox{}^* d g + g^2 = [1-S] g\, \mbox{}^* d f + g^2 + \mbox{Total divergence.}.
\ee
Substituting by (\ref{semi1}), we have finally:
\be
L_p(F) = [2S-1] (df)^2 + \mbox{Total divergence}.\label{mod1}
\ee
On the other hand, vary now with respect to $f$:

\be
\mbox{}^* d g=0; \label{semi3}
\ee
as consequence of this, there exists ${\bar g}$ such that \be g=d{\bar g}\label{semi4}.\ee
Integrating once by parts (in the opposite sense) :
\be
L_p = g\, \mbox{}^* d f - S f\, \mbox{}^* d g + g^2 = [S-1] f\, \mbox{}^* d g + g^2 + \mbox{Total divergence.}.
\ee
One can express the master action as a function of ${\bar g}$; substituting
 here by the equation of motion (\ref{semi3}) and (\ref{semi4}), we obtain:
\be
L_p = (d{\bar g})^2 + \mbox{Total divergence.}.\label{mod2}
\ee
This proves the duality between two Maxwell-type models, (\ref{mod1}) and (\ref{mod2}),
obtaining, on-shell, the two dual actions:
\be
L_p(F(f))= [2S-1] [df]^2
\ee
and
\be
L_p (F({\bar g}))= [d{\bar g}]^2
\ee

If we take the $k_g$ to be zero (and not $k_f$),
the result is another duality between two Maxwell-type models
for the fields $g$ and ${\bar f}$, where $f=d{\bar f}$

\subsection{Dual Equivalence between Proca models.}

Consider our master action given by (\ref{GSDM}), with non-singular 
matrix $K$.
 By taking the variation with respect to $f$, we
obtain: \be \label{pr1}\mbox{}^* df = k_g g , \ee

thus, applying $\mbox{}^*$ to both sides of this equation, we get

\be df = S k_g \, \mbox{}^* g. \label{semi2}\ee

Substituting (\ref{pr1}) and (\ref{semi2}) in the action, the final result is
\be L(f,g[f])= \frac{ 2S-1}{k_g} df^2 + k_f f^2 .\ee

 Variation with respect to $g$ results in
 \be
 L(g,f[g])= \frac{ 2S-1}{k_f} dg^2 + k_g^2 g^2 .
 \ee
This constitutes the proof that the dual equivalence between two
 Proca theories is described by the action
(\ref{GSDM}).
\section{(2+1)-dimensions: Maxwell Theory and manifest HD.}

In order to show one of the numerous possible applications of the doublets structures,
we briefly describe HD
    in 3d, which would be the 3d-version
   of a electric-magnetic duality. This has also been discussed in
   another contexts \cite{RW,ban-w}.

As we have pointed in relation to equation (\ref{a10}) in the
Introduction, HD in (2+1)-dimensions is ill defined unless we use
doublets of different tensor rank as we show below; otherwise, we
only can work merely with the already known DHD described in the
introduction by the action (\ref{180}).

Let us consider the Maxwell-type theory in three dimensions:

\be
 S=\int dx^3 {\cal F}^2 \, \label{eq0}
\ee
where ${\cal F}$ is a doublet,
\be
{\cal F}=(F_{\mu\nu};f_\rho),
\ee
where ${\cal F}=d{\cal A}$ and ${\cal A}=(A_\mu ; a)$.

Notice that this is the {\it unique} doublet that can be chosen such that
its components are a differential of some potential-field.

Now, the self-duality (HD) is well-defined \footnote{In (2+1)-dimensions the metric has two minuses,
 thus: $S=1$.}: \be {\cal F}= \mbox{}^* {\cal F}, \ee this means the
simultaneous relations: \be f_\rho = \ep_{\mu\nu\rho} F^{\mu\nu},
\label{eq3}\ee \be F_{\mu\nu} =  \ep_{\mu\nu\rho} f^\rho .\label{eq4}\ee

The Electric and Magnetic part, for both $f,F$, can be defined in the usual way and the relations   
(\ref{eq3}) and (\ref{eq4}) relate each other.

The equations of motion for (\ref{eq0}) read
\be
\label{eqcalF}
div {\cal F}=0.
\ee

As a consequence of the HD-relations (\ref{eq3}) and (\ref{eq4}),
one of these (two) equations reduces to be an {\it identity} (the Bianchi identity).
However, these HD-relations do not arise from the action as consequence of the equations
 of motion.

In order to render this duality manifest, we again can use our GSDM and redefine the doublet
 structure to
 be potential-field type. Let us take
\be
{\cal B}=(a, F_{\mu\nu})
\ee
we shall show that this theory can be described by a semiself-dual action (one more time),
 including
self-duality relations. The semi-self-dual action proposed is
\be
S=\int d^3 x \left({\cal B}\, .\mbox{}^{*}d   {\cal B} + F_{\mu\nu}F^{\mu\nu}\right) . \label{eq5}
\ee
The equations of motion are

\be F_{\mu\nu} =  \ep_{\mu\nu\rho} \partial^\rho a \label{eq1}\ee
and
\be  \ep_{\mu\nu\rho} \partial^\rho F^{\mu\nu}=0.
\ee
This implies that $ div F=0$ and that there exists $A_{\mu}$ such that
 $F_{\mu\nu}=\partial_{[\mu}A_{\nu]}$, thus, defining $f_\mu=\partial_{\mu} a$,
we obtain from (\ref{eq1}):
\be
\label{HD2}
f_\rho = \ep_{\mu\nu\rho} \partial^\nu A^\rho,
\ee
and then $div f=0$. Equations of motion (\ref{eqcalF}) are verified and
HD-relations are recovered (Equations (\ref{HD2}) and (\ref{eq1}).

Finally, notice that in four dimensions it is more clearer to build this up, since
 it is possible construct
 a doublet of Maxwell field strengths and the corresponding Hodge type self-duality (HD)
 \cite{SS,Z,DT}; and in contrast, {\it DHD can be defined} in four dimensions too. 
 Realizations in 4d with manifest DHD are precisely the actions $S_{\cal F}$
 and $S_{\cal G}$ (with $K$ generic) discussed in Section 3.1.
In \cite{ban95,t1}, the equivalence of these models to Proca
 and massive Kalb-Ramond theories is discussed.

\section{Concluding remarks.}

 In this work, we have extended the notion of duality for all type of tensors
  in arbitrary dimensions in order to allow, a well-defined notion of
   self-duality. We have also built a general action describing this fact, which
   is shown to give (by taking proper limits in its parameters) the parent actions in the most of the case of dual models.
Interesting possibilities that we open up as an application of the results presented here 
shall be explored:
this action (GSDM) is related
\cite{t1} to the topologically massive, the so-called BF-theories \cite{BF}, via the manifest 
connection with the SD-model in three dimensions \cite{tmdob}
presented here; furthermore,  the
bosonization technique in arbitrary dimensions, mainly in higher dimensions ($d\geq 4$) ,
thanks to this connection,
comes out related to a topologically massive model that mixes different gauge forms. These results
 shall soon be reported \cite{boson}.

that we open up as an application of the results presented here 
is the study of bosonization in arbitrary dimensions, mainly in higher dimensions. This is
 not a trivial matter \cite{luscher,marino,banmarino}, but with the help
 of the technique suggested here, $d\geq 4$ bosonization comes out 
in connection with a topologically massive model that mixes different gauge forms. Results
 on this issue shall
soon be reported elsewhere \cite{bos}. 

, which found important applications\cite{BF}.

A novel definition of HD, and non merely the
    already known DHD, has been given in (2+1)-dimensions. Reciprocally, the approach presented here,
	allows us write DHD in even dimensional spacetime.

   There are several potential applications of this structure, for saying
   supersymmetric
 extensions of
these general self-dual actions. Other approaches on this matter have alrady been made in the past\cite{kn:GHR,kn:A}.

Furthermore, as has been pointed in \cite{con1,con2,con3}, a 
self-duality relating massive
 scalar and vector fields may be relevant for
string theory in the context of massive type IIA Sugra, however
 the self-duality presented here is new.

  We suggest, finally, a hypothesis motivated in all these examples:
  " All parent action interpolating between two dual models comes from
  (in the sense discussed in this work) a
  GSDM"; in other words, every duality at the level of the classical actions
  comes from some manifest duality between the fields involved in these actions.

{\bf Aknowledgements}: The author is indebted to Clovis Wotzasek
for suggesting this interesting research line and for many invaluable discussions.
  Special thanks are due also to J. A. Helayel-Neto. CNPq is
acknowledged for the invaluable financial help.


\begin{thebibliography}{99}

\bib {PST1} P. Pasti, D. Sorokin and M. Tonin, Phys. Lett. B352 (1995) 59;
Phys. Rev. D52 (1995) R4277; S. Deser, Lectures given at 7th Mexican School
of Particles and Fields and 1st Latin American Symposium on High-Energy Physics,
Merida, Yucatan, Mexico, 1996, hep-th/9701157;
D.I. Olive, Introduction to duality,
In *Cambridge 1997, Duality and supersymmetric theories* 62-94.



\bibitem{review} {O} D.I. Olive, Exact electromagnetic duality,  Nucl.
Phys. B(Proc. Suppl.) 58 (1997) 43, hep-th/9508089; {GH} C. Gomez
and R. Hernandez, Electric-magnetic duality and effective field
theories, hep-th/9510023;  J. Schwarz, Lectures on superstring and
M theory dualities, Nucl. Phys. Proc. Suppl. 55B (1997) 1,
hep-th/9607201.

\bibitem{witten}E. Witten, ELECTRIC MAGNETIC DUALITY IN FOUR-DIMENSIONAL GAUGE THEORIES,
11th International Conference on Mathematical Physics: New
Problems in the General Theory of Fields and Particles, Paris,
France, 25-28 Jul 1994.

\bibitem{TPvN} P. K. Townsend, K. Pilch and P. van Nieuwenhuizen, Phys.
Lett. B 136 (1984) 38.
\bibitem{annals} R.Jackiw, S. Deser and Templeton, Ann. Phys. 140 (1982) 372.

\bibitem{DJ} S. Deser and R. Jackiw, Phys. Lett. B 139 (1984) 2366.

\bibitem{kn:DGHT} S. Deser, A. Gomberoff, M Henneaux and C. Teitelboim, {\it Duality, Self-Duality, Sources and Charge Quantization in Abelian $N$-Form Theories}, hep-th/{\bf 9702184};

\bibitem{W}C. Wotzasek, Phys.Rev.D58:125026,1998

\bibitem{ban-w}Rabin Banerjee, Clovis Wotzasek, Phys.Rev.D63:045005,2001

\bib{EK} J. Schwarz, Lectures on superstring and M theory dualities,
Nucl. Phys. Proc. Suppl. 55B (1997) 1;
Elias Kiritsis,
Lectures given at NATO Advanced Study Institute: TMR Summer School
on Progress in String Theory and M-Theory , Cargese,
hep-ph/9911525--------

\bibitem{tmdob} M. Botta Cantcheff, ''
Topologically massive gauge theories from first order theories in
arbitrary dimensions ", hep-th/0110264.

\bibitem{bos}  M. Botta Cantcheff, J. A. Helayel-Neto, work in progress.

\bibitem{suecos} For a review in the use of the master action
 in proving duality in diverse areas see: S. E. Hjelmeland, U. Lindstr\"om, UIO-PHYS-97-03, May 1997.
e-Print Archive: hep-th/9705122 and references therein.

\bibitem{ban95} Banerjee, N. and Banerjee, R., Mod. Phys. Lett. A11 (1996) 1919.


\bib {SS} J. Schwarz and A. Sen, Nucl. Phys. B411 (1994) 35

\bib {Z} D. Zwanziger, Phys. Rev. D3 (1971) 880
\bib {DT} S. Deser and C. Teitelboim, Phys. Rev. D13 (1976) 1592.
\bib{EK} J. Schwarz, Lectures on superstring and M theory dualities,
Nucl. Phys. Proc. Suppl. 55B (1997) 1;
Elias Kiritsis,
Lectures given at NATO Advanced Study Institute: TMR Summer School
on Progress in String Theory and M-Theory , Cargese,
hep-ph/9911525--------


\bibitem{RW}R.Banerjee and C.Wotzasek, Nucl.Phys B527(1998) 402

\bibitem{BF} A. Aurilia, Y. Takahashi, Prog.Theor.Phys.66:693,1981;
T. J. Allen, M. J. Bowick, A. Lahiri, Mod.Phys.Lett.A6:559-572,1991 

\bibitem{t1}
E. Harikumar, M. sivakumar, Nucl Phys -B565(2000), 385; Mod.Phys Lett A 15 (2000) 121.

\bibitem{kn:GHR} S. J. Gates Jr, C. M. Hull and M. Ro\v{c}ek,
 {\it Twisted Multiplets and New Supersymmetric Non-Linear Sigma Models}, Nucl. Phys. {\bf B248} (1984) 157;

\bibitem{kn:A} M. Ro\v{c}ek, K. Schoutens and A. Sevrin,
 {\it Off-shell WZW Models in Extended Superspace}, Phys. Lett. {\bf B265} (1991) 303;

\bibitem{con1}H. Gustafsson, S.E. Hjelmeland, U. Lindstrom, Phys.Scripta 60:305-311,1999.

\bibitem{con2}L.J. Romans, Phys.Lett.B169:374,1986.

\bibitem{con3} Bergshoeff, M. de Roo, M.B. Green, G. Papadopoulos, P.K. Townsend, Nucl.Phys.B470:113-135,1996.

   
 
\end{thebibliography}
\end{document}